\documentclass[letterpaper, 10 pt, conference]{ieeeconf}  

\IEEEoverridecommandlockouts                              

\makeatletter

\let\proof\@undefined
\let\endproof\@undefined
\makeatother

\usepackage{amsmath, lmodern, bm} 
\usepackage{amssymb} 
\usepackage{amsthm}
\usepackage{dsfont}
\usepackage{siunitx} 

\usepackage{color, colortbl}
\usepackage[bottom]{footmisc}

\usepackage{float}
\usepackage{subfigure}
\usepackage{graphicx}
\usepackage{caption}
\usepackage[ruled]{algorithm2e}

\usepackage{cite}

\usepackage{tikz}
\usetikzlibrary{shapes,arrows,positioning,calc,backgrounds}

\newcommand{\vect}[1]{\mathbf{#1}}

\DeclareMathOperator*{\argmax}{arg\,max}

\definecolor{gray}{rgb}{0.5, 0.5, 0.5}

\definecolor{gray}{rgb}{0.5, 0.5, 0.5}
\definecolor{lightgray}{rgb}{0.83, 0.83, 0.83}

\newtheorem*{problem}{Problem}

\usepackage[normalem]{ulem}

\title{\LARGE \bf
Online and Adaptive Parking Availability Mapping:\\ An Uncertainty-Aware Active Sensing Approach\\ for Connected Vehicles}

\author{Luca~Varotto
        and~Angelo~Cenedese%
\thanks{L. Varotto and A. Cenedese are with the Department of Information Engineering, University of Padova, Italy.
A.~Cenedese is also with the Institute of Electronics, Information and Telecommunication Engineering, National Research Council (CNR - IEIIT).
Corresponding author: {\tt\small luca.varotto.5@phd.unipd.it}.}
\thanks{This work was partially supported by the Department of Information Engineering under the BIRD-SEED TSTARK project.}
}%

\begin{document}

\maketitle
\thispagestyle{empty}
\pagestyle{empty}

\begin{abstract}
Research on connected vehicles represents
a continuously evolving technological domain, fostered by the emerging Internet of Things (IoT) paradigm and the recent advances in intelligent transportation systems. Nowadays, vehicles are platforms capable of generating, receiving and automatically act based on large amount of data.
In the context of assisted driving, connected vehicle technology provides real-time information  about the  surrounding  traffic  conditions. Such information is expected to improve drivers’ quality of life, for example, by adopting decision making strategies according to the current parking availability status.
In this context, we propose
an online and adaptive scheme for parking availability mapping. Specifically, we adopt an information-seeking active sensing approach to select the incoming data, thus preserving the onboard storage and processing resources; then, we estimate the parking availability through Gaussian Process Regression.
We compare the proposed algorithm with several baselines, which attain inferior performance in terms of mapping convergence speed and adaptivity capabilities; moreover, the proposed approach comes at the cost of a very small computational demand.
\end{abstract}

\section{Introduction}\label{sec:intro}

\begin{figure*}[h!]
\centering
\includegraphics[width=0.55\textwidth
]{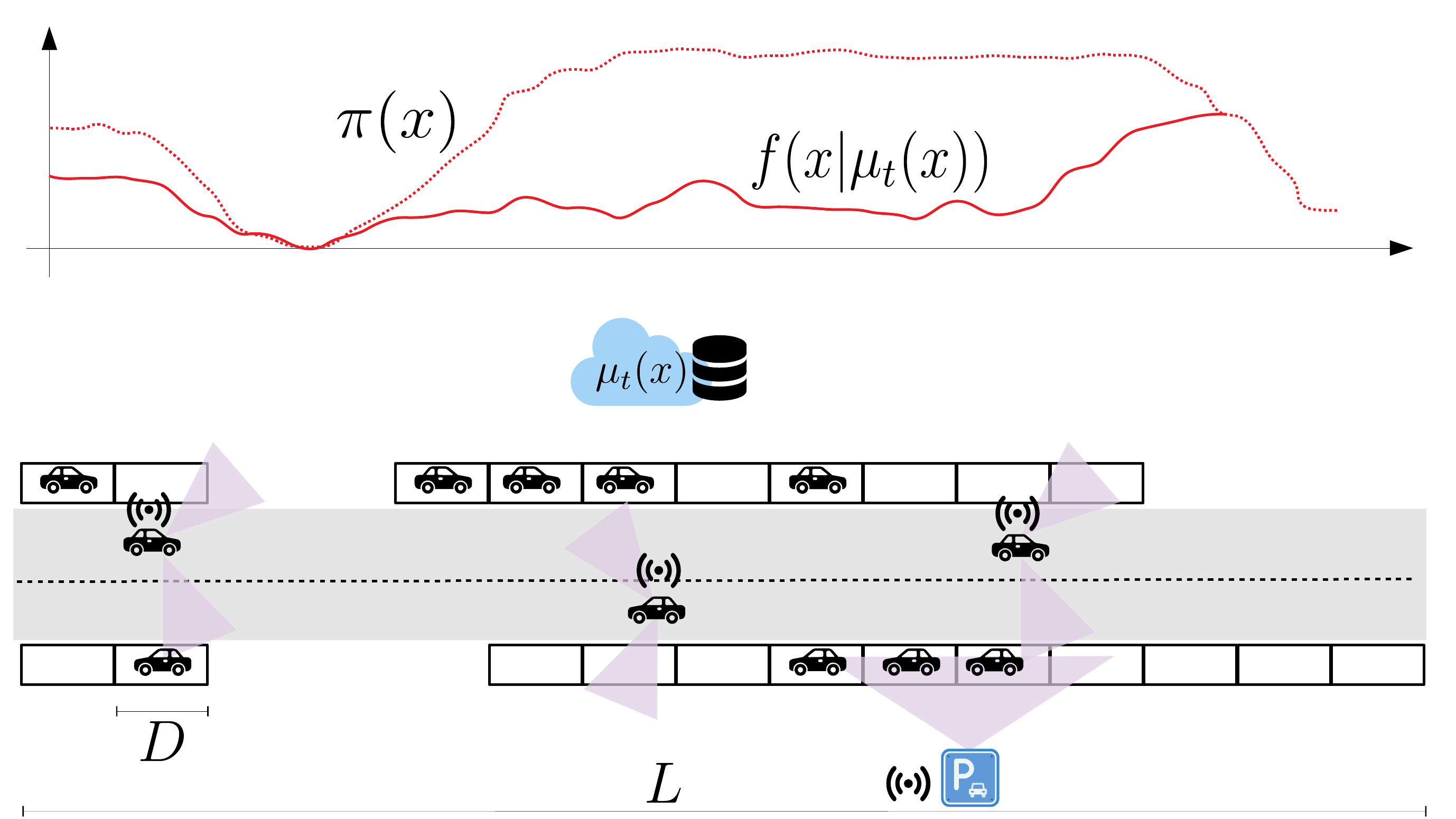}
\caption{Problem scenario. A road path of length $L$ is populated by intelligent and connected vehicles, as well as with smart parking infrastructures. Each vehicle is capable to detect parking slots and to classify them as free or occupied. Traffic conditions are grabbed from cloud databases; hence, by establishing communication with external information sources (other vehicles, parking stations), each vehicle can learn the parking availability map over the entire road path.}
\vspace{-0.5cm}
\label{fig:scenario}
\end{figure*}

The research on Advanced Driver Assistance Systems (ADASs) is  an extremely active and promising field of study, across all levels from the Level~1 of \emph{Driver Assistance}, with possible intelligent intervention to steering, braking, and acceleration, to the futuristic Level~5 of \emph{Full Automation}, where the vehicle can operate all driving conditions~\cite{sae2014taxonomy}.
Along these avenues, though, it has been recently highlighted how to unleash the full potential of autonomous driving, the smartness of the infrastructure is at least as important as the intelligence inside the vehicle itself~\cite{joy2017internet, gopalswamy2018infrastructure}, where by \emph{infrastructure} we mean both the information and communication systems distributed along the roads that will support the autonomous driving capabilities and the dynamic information-rich collaborating network that emerges from the many vehicles that are actually circulating~\cite{uhlemann2015introducing,uhlemann2018time}. 

Such an information flow allows the definition of real-time maps that are needed for efficient and safe navigation: to this aim, mapping is an all-inclusive procedure that accounts for topology of roads, geometry of the environment, and more volatile data such as, for example, traffic levels, congestion or accident occurrences, parking availability~\cite{wong2021mapping}.
In order to acquire such mixture of permanent and temporary information, active sensing methodologies come into play, where the information acquisition process is controlled and driven by some optimal criteria chosen according to the task or the goal to be accomplished.
%
%
%
Specifically, the vehicle can be interpreted as a mobile sensing platform that leverages environmental perception from on-board sensors, information coming from fixed infrastructure, and data exchange with other vehicles to perform scene reconstruction and understanding~\cite{siegel2017survey}. 
On the other hand, the combination with active sensing approaches allows to improve the quality of information and the efficiency of the information gathering process~\cite{varotto2021probabilistic,varotto2021transmitter}.
This framework results particularly effective when the information of interest is highly dynamic, such as that related to traffic condition and parking availability. In this case, the data coming from the network of connected vehicles provide high-fidelity real-time information that supports the driver's decision making process and improves the overall system mobility~\cite{talebpour2016influence}. 


With such premises, the parking forecast problem can be seen as a probabilistic map building problem that can be addressed through an incremental information discovery process: namely, the problem is to estimate the parking availability in a certain objective area by gathering information along the path towards the desired place, and to take actions accordingly in order to maximize the parking probability itself.








\noindent \textbf{Related works -}
%
%
In the last few years, there has been a growing literature on the design of strategies for smart parking, and on the development of suitable models for parking, thanks to the strong demand for ADAS technologies and the quest for the autonomous vehicle (see~\cite{lin2017survey,bischoff2018autonomous} and references within).

Indeed, smart vehicles equipped with assistive systems exploit semantic maps of the environment to navigate and operate. In~\cite{westfechtel2018parking}, for example, a map of the environments is build from on-board sensors (basically LiDAR and GPS) and a graph-based approach is proposed to include topological and usage parking information, obtained also from multiple days observation data. This map is then used to evaluate the parking spots occupancy rate and therefore predict their availability. 
%
%
%
A collaborative approach is proposed in~\cite{li2018collaborative} with the specific focus of the management of parking garages, where a global probabilistic occupancy map emerges from individual vehicle perception (e.g. through LiDAR, IMU, GPS) and inter-vehicle information sharing to support autonomous parking from the available parking spot identification, to the navigation and motion planning, to the precise parking maneuvering. 
%
%
The multi-vehicle data collection is considered also in the ParkNet system~\cite{mathur2010parknet}, where the problem of urban street parking is addressed by means of drive-by parking monitoring; the collected data from GPS and ultrasonic rangefinders are then fused into a single centralized map, which allows the generation of statistics for optimal management policies.
%
%
%

On the modeling aspects, also, the state of the art reports several research avenues. 
In particular, the employment of Gaussian Processes (GPs) to model occupancy maps appears as a convenient method to attain a probabilistic description of the behavior of mobile objects in the environments, given their physical spatial structure and correlation~\cite{o2009contextual}.
%
%
%
Dynamic GP occupancy maps are further developed in~\cite{senanayake2017learning}, where variational inference methods are employed 
to speed up the estimation process and adapt to LiDAR produced dataflow.
%
%
%
A different approach is taken by other models that instead of focusing on the occupancy map, consider the point of view of the vehicles that are moving towards the parking slots. For example,   in~\cite{rajabioun2015street}, a multivariate autoregressive model is proposed where temporal and spatial correlations are taken into account to compute the probability of parking availability at the vehicle estimated arrival time; this system exploits real-time and historical data, where the latter are used to learn the parameters of the model and the former are used to carry out an accurate up-to-date prediction.
%
%
%
Finally, queuing models are a popular and established method to describe the behavior of the vehicles-parking slots interaction and are employed in~\cite{tavafoghi2019queuing} to provide probabilistic estimation of parking availability, leveraging on learning techniques to attain the prediction model parameters.

\noindent \textbf{Contributions -} 
This paper considers a smart transportation scenario, populated by smart infrastructures (I), as well as intelligent and connected vehicles (V)~\cite{uhlemann2018time}. These are able to detect parking slots and to classify them as either free or occupied~\cite{westfechtel2018parking}.
They are also endowed with communication capabilities, so that V2V and V2I data exchange are allowed.  

In this framework, the purpose is the design of an online parking mapping scheme, which leverages on local and remote data to adapt to dynamic traffic conditions. To this aim, we formulate the task as a Gaussian Process Regression (GPR) problem~\cite{williams2006gaussian}, coupled with an active sensing module; this is based on an information-seeking data selection criterion, which retains only those datapoints that are expected to reduce the uncertainty of the model. 
%
To the best of authors' knowledge, this is the first attempt to design a parking mapping algorithm by leveraging active sensing within the framework of connected vehicles. 
In particular, with respect to the reviewed literature, the proposed methodology employs a data-driven approach (i.e., GPR) to create the parking availability map; to this aim, the training dataset is collected in a multi-source scenario, where V2V and V2I communication guarantees incremental learning and continuous map update. Finally, the active sensing module selects the most informative samples, thus avoiding memory and processing overload, typical of data-driven multi-source learning procedures.
Numerical results validate the effectiveness of the proposed algorithm against several baselines: the combination of GPR with active sensing is shown to improve the quality and the efficiency of the estimation process. In particular, the suggested approach  attains superior performance in terms of mapping convergence speed and adaptivity capabilities, at the cost of small computational demands.

\section{Problem statement}\label{sec:problem_formulation}

Fig. \ref{fig:scenario} provides an overview of the problem scenario: 
a road path is populated by mobile sensing platforms (e.g., autonomous vehicle~\cite{bischoff2018autonomous})
with sensing, processing and communication capabilities onboard (i.e., intelligent and connected vehicles~\cite{uhlemann2018time,siegel2017survey}). Each platform establishes communication with surrounding smart infrastructures (e.g., smart parking stations) and with the other vehicles, according to the V2I~\cite{uhlemann2015introducing} and V2V~\cite{krasniqi2016use} paradigms. The platform has also access to cloud storage resources, where information on current traffic conditions can be downloaded. 
The purpose is to gather, from onboard sensors or from external resources, informative data to estimate the parking availability over the entire route\footnote{Bold letters indicate (column) vectors, if lowercase, matrices otherwise. $\vect{I}$ is the identity matrix.
Regarding the statistical distributions used in this paper, $\mathcal{N}(x|\mu,\sigma^2)$ is the Gaussian distribution over the random variable $x$ with expectation $\mu$ and variance $\sigma^2$; a Bernoulli distribution with parameter $p$ is denoted as $\mathcal{B}(p)$; a uniform distribution in $[a,b]$ is denoted as $\mathcal{U}[a,b]$.
}.  

\subsection{Environment}\label{subsec:environment}
We consider a road path $\mathcal{X} =[0,L]$ over which the following function is defined
\begin{equation}\label{eq:prior}
    \pi(x): \mathcal{X} \mapsto [0,1].
\end{equation}
It represents the (on-street~\cite{rajabioun2015street,mathur2010parknet}) \emph{a-priori parking availability} along the path $\mathcal{X}$. Being related to structural and environmental properties of the street in traffic-free conditions (e.g., presence and number of parking slots), it is a time-invariant quantity. In this work, $\pi(x)$ is the total count of slots over a spatial window $W$, namely
\begin{equation}\label{eq:cumulative_prior}
    \pi(x) = \frac{1}{\lfloor W/D\rfloor} \sum_{h=0}^{\lfloor W/D\rfloor} \mathds{1}(x-hD) , \quad x \in \mathcal{X}
\end{equation}
where $D$ is the length of a parking slot (homogeneous slots are considered); $\lfloor W/D\rfloor$ is the maximum number of parking slots over the window $W$ and it works as normalization factor; $\mathds{1}(\cdot)$ is an indicator function, such that \mbox{$\mathds{1}(x-hD)=1$} if between \mbox{$x-hD$ and $x-(h+1)D$} there is a parking slot; \mbox{$\mathds{1}(x-hD)=0$} otherwise. 
In practice, $\pi(x)$ is produced once and for all via parking spot detection and mapping algorithms~\cite{westfechtel2018parking}. 

The road path is supposed to be characterized, at each time instant $t$, by a time-varying \emph{traffic density} function
\begin{equation}\label{eq:traffic_density}
    \mu_t(x): \mathcal{X} \mapsto [0,1].
\end{equation}
This function maps each point of $\mathcal{X}$ into a normalized density value, proportional to the traffic congestion level. We then suppose that there exists a \emph{parking availability attenuation function}, which depends on the current traffic density level~\cite{aryandoust2019city,bischoff2018autonomous,cao2015system}, namely
\begin{equation}\label{eq:attenuation}
    \lambda\left(\mu_t(x)\right): [0,1] \mapsto [0,1].
\end{equation}
This function defines the parking availability reduction according to the traffic density $\mu_t(x)$ at a certain location $x$. Hence, the \emph{parking availability map} (PAM) (also referred to as occupancy map~\cite{westfechtel2018parking}) is computed as 
\begin{align}\label{eq:PA_true}
    f\left(x|\mu_t(x)\right): \;
    \mathcal{X} & \mapsto [0,1] \\ 
    x & \rightarrow  \pi(x)\lambda\left(\mu_t(x)\right),
    \end{align}
where notation $f\left(x|\mu_t(x)\right)$ means that $f(x)$ is computed, given $\mu_t(x)$ (for ease of notation, $f\left(x|\mu_t(x)\right)$ is sometimes referred also as $f\left(x\right)$). By aggregating information on environmental properties (through $\pi(x)$) and on current traffic conditions (through $\mu_t(x)$), the PAM represents the number of available parking slots between $x$ and $x-W$, normalized by the maximum number of parking slots (i.e., $\lfloor W/D\rfloor$). In fact, it is computed as 
\begin{equation}\label{eq:PA_indicator}
    f\left(x|\mu_t(x)\right) 
    = \frac{1}{\lfloor W/D\rfloor} \sum_{h=0}^{\lfloor W/D\rfloor} \mathds{1}_{\mu_t}(x-hD) , \quad x \in \mathcal{X}
\end{equation}
where $\mathds{1}_{\mu_t}(x-hD)$ is an indicator function, such that, at time $t$, \mbox{$\mathds{1}_{\mu_t}(x-hD)=1$} if between \mbox{$x-hD$ and $x-(h+1)D$} there is a parking slot and it is available; \mbox{$\mathds{1}_{\mu_t}(x-hD)=0$} otherwise.

\subsection{Sensing platform}\label{subsec:platform}
This work leverages on the active sensing framework~\cite{varotto2021wakeUp} to solve the parking availability estimation problem~\cite{mathur2010parknet}; therefore, the main element of the problem scenario is the sensing platform, which interacts with the surrounding environment and actively selects the information to gather from it.  

\subsubsection{\textbf{Platform observation model}}\label{subsubsec:observation_model}
We consider a mobile sensing platform equipped with communication capabilities (i.e., \emph{connected vehicle}~\cite{uhlemann2018time}). The platform  embeds also sensing, computational and logical resources onboard, so that it is capable to detect on-street parking slots and to recognize their availability, as in~\cite{bischoff2018autonomous}. Thus, we consider the following platform measurement
\begin{equation}\label{eq:PA_meas}
    y_t(s_t) = \frac{1}{\lfloor W/D\rfloor} \sum_{h=0}^{\lfloor W/D\rfloor} \tilde{\mathds{1}}_{\mu_t}(s_t-hD) 
\end{equation}
where $s_t \in \mathcal{X}$ is the platform state at time $t$, represented by the vehicle location over the road path. The binary function
$\tilde{\mathds{1}}_{\mu_t}(s_t-hD) \in \{0,1\}$ is the \textit{estimated occupancy}: $\tilde{\mathds{1}}_{\mu_t}(s_t-hD)=1$ if the platform classifies the slot between $s_t-hD$ and $s_t-(h+1)D$ as available; \mbox{$\tilde{\mathds{1}}_{\mu_t}(s_t-hD)=0$} otherwise. To account for sensors nuisance and for the estimation uncertainty associated to common parking slot classification algorithms~\cite{bischoff2018autonomous}, $\tilde{\mathds{1}}_{\mu_t}(s_t-hD)$ is a noisy version of the true value \mbox{$\mathds{1}_{\mu_t}(s_t-hD)$}. In particular, we consider the following \emph{PA observation model}
\begin{align}\label{eq:observation_model}
    y_t(s_t)
    & = \underbrace{\frac{1}{\lfloor W/D\rfloor} \sum_{h=0}^{\lfloor W/D\rfloor} \mathds{1}_{\mu_t}(s_t-hD)(1+\omega_1)}_{y_{t,1}} \\
    & + \underbrace{\frac{1}{\lfloor W/D\rfloor} \sum_{h=0}^{\lfloor W/D\rfloor} \left(1-\mathds{1}_{\mu_t}(s_t-hD)\right)\omega_0}_{y_{t,0}} \nonumber
\end{align}
where $\omega_0$ and $\omega_1$ are noisy terms: \mbox{$\omega_1 \sim \mathcal{B}(p_1)$} with support in $\{-1,0\}$, while \mbox{$\omega_0 \sim \mathcal{B}(p_0)$} with support in $\{0,1\}$. In practice, $\omega_1$ induces a free slot to be either not detected or detected as occupied; hence, it accounts for mis-detections and mis-classifications. On the other hand, $\omega_0$ induces an absent or occupied slot to be seen as a free parking space; thus, it accounts for sensor artifacts and mis-classifications as well.
From the Bernoulli distribution of $\omega_0$ and $\omega_1$, it follows that $y_{t,0}$ and $y_{t,1}$ are Binomial random variables; by exploiting the Gaussian approximation of Binomial random variables, we get the following approximation of \eqref{eq:observation_model}
\begin{equation}\label{eq:observation_model_gaussian}
\begin{split}
    & y_t(s_t) \approx f(s_t|\mu_t(s_t)) + \epsilon_t\\
    & \epsilon_t \sim \mathcal{N}\left(\epsilon \left \vert{0, \sigma^2}\right. \right).
\end{split}
\end{equation}
Note that the observation model \eqref{eq:observation_model_gaussian} is linear, with noise component concentrated in the Gaussian term $\epsilon_t$.

\subsubsection{\textbf{External information sources}}\label{subsubsec:connected}

In line with the IoT paradigm, the platform is supposed to receive information from \emph{external sources}, such as other connected vehicles, cloud databases, and smart parking stations~\cite{krasniqi2016use,uhlemann2018time}. The set of external sources that establish communication with the platform at time $t$ is denoted as 
$
\mathcal{V}_t = \{ v_{t,i} \}_{i=1}^{N_t}
$, while the location of the $i$-th source is indicated with $x_i \in \mathcal{X}$. 
We suppose that each agent in $\mathcal{V}_t$ sends to the platform the measured PAM at its location, namely  
$
y_t( x_i), \; i=1,\dots,N_t,
$
where each $y_t( x_i)$ follows \eqref{eq:observation_model_gaussian}.
We also assume that the platform has access to an external database from which perfect information on the traffic density level over the entire path can be gathered; in other words, at each time instant $t$, the platform knows $\mu_t(x), \; \forall x \in \mathcal{X}$.

\subsubsection{\textbf{The PAME problem}}\label{subsubsec:problem}
Given the premises provided in Sec. \ref{subsubsec:observation_model} and \ref{subsubsec:connected}, the platform can accumulate large amount of data over time; in particular, if no data selection strategies are applied, at time $t$ the platform stores the following dataset  
\begin{equation}\label{eq:dataset}
    \mathcal{D}_t = \left( \vect{X}_t,\vect{y}_t \right).
\end{equation}
This is incrementally fed with new incoming data, namely
\begin{equation}\label{eq:dataset_update}
\begin{split}
    & \vect{X}_t = \vect{X}_{t-T} \cup s_t \cup \{ x_i \}_{i=1}^{N_t} \\
    & \vect{y}_t = \vect{y}_{t-T} \cup y_t( s_t) \cup \{ y_t( x_i) \}_{i=1}^{N_t}.
\end{split}
\end{equation}
where $T$ is the sampling time according to which the platform collects a new measurement, updates the knowledge on $\mu_t(x)$, and processes the accumulated data received from the external sources since the last iteration at $t-T$; in other words, the set of data $\{ x_i \}_{i=1}^{N_t}$ is accumulated between $t_T$ and $t$ and processed at $t$.

Notably, the cardinality of $\mathcal{D}_t$ is
\begin{equation}\label{eq:dataset_cardinality}
    |\mathcal{D}_t| = t + \sum_{\tau=1}^t N_\tau,
\end{equation}
which linearly increases as the time grows. Consequently, to operate with the limited onboard computational and storage resources of the platform, efficient data collection, filtering and management schemes are required (e.g., active sensing algorithms~\cite{varotto2021wakeUp}). 

\begin{problem} With the formalism introduced in this Section, the \emph{parking availability map estimation} problem (PAME) can be formulated as the reconstruction of the latent PAM function $f(x|\mu_t(x))$, over $\mathcal{X}$ and with data in $\mathcal{D}_t$. 
The problem is solved \emph{online} when the reconstructed function, $\hat{f}(x|\mu_t(x))$, is updated as new samples are collected. 
A PAME technique is \emph{adaptive} if it can handle the time-varying fluctuations of the traffic density level, namely $\mu_t(x) \neq \mu_{t-T}(x)$.
\end{problem}

\section{Methodology}\label{sec:method}

To solve the PAME problem defined in Sec. \ref{sec:problem_formulation}, we propose a GPR scheme that incrementally and adaptively learns the underlying function $f(x|\mu_t(x))$, leveraging on $\mathcal{D}_t$. As GPR strategies do not scale well with the cardinality of the dataset~\cite{das2018fast,liu2020gaussian,moore2016fast,shahriari2015taking}, we also propose an active sensing method to select the incoming samples, according to an uncertainty-aware acquisition policy. 

\subsection{GPR-based PAME}\label{subsec:PAME_GPR}

\subsubsection{\textbf{GPR}}\label{subsubsec:GPR}

A Gaussian Process (GP) is a collection of random variables, any finite number of which have a joint Gaussian distribution~\cite{williams2006gaussian}. Given the input vector \mbox{$\vect{x} \in \mathbb{R}^p$}, a GP is fully specified by its mean function $m(\vect{x})$ and covariance function $k(\vect{x},\vect{x}^\prime)$, namely
\begin{equation}\label{eq:GP_prior}
    f(\vect{x}) \sim \mathcal{GP}\left(m(\vect{x}),k(\vect{x},\vect{x}^\prime)\right),
\end{equation}
where $f:\ \mathbb{R}^p \rightarrow \mathbb{R}$ is referred to as {\em latent function}. 
Mean and covariance (or kernel) functions incorporate prior knowledge (e.g., periodicity, smoothness) about the latent function.
The mean function is typically constant (either zero or the mean of the training dataset), while the most commonly-used kernel functions are constant, linear, square exponential or Matern, as well as compositions of multiple kernels~\cite{williams2006gaussian}.
The hyperparameters in the mean and covariance functions are computed by fitting the train dataset $\mathcal{D}$ of cardinality $n_{train}$ 
\begin{equation}\label{eq:dataset}
    \begin{split}
& \mathcal{D} = \left\lbrace (\vect{x}_i,y_i)  \right\rbrace_{i=1}^{n_{train}} = (\vect{X},\vect{y}) \\
    & \vect{X} = 
\begin{bmatrix}
\vect{x}_1 & \dots & \vect{x}_{n_{train}}
\end{bmatrix}^\top \in \mathbb{R}^{n_{train} \times p} \\
    & \vect{y} =
\begin{bmatrix}
y_1 & \dots & y_{n_{train}}
\end{bmatrix}^\top
\in \mathbb{R}^{n_{train}}.
    \end{split}
\end{equation}
Each training label $y_i$ is a noisy measurement of the latent function $f(\vect{x})$, namely
\begin{equation}\label{eq:GP_noisy_labels}
    y_i = f(\vect{x}_i) + \epsilon_i, \quad \epsilon_i \sim \mathcal{N}(\epsilon|0,\sigma^2).
\end{equation}
To account for the i.i.d. Gaussian noise in the training labels, the GP in \eqref{eq:GP_prior} becomes~\cite{williams2006gaussian},
\begin{equation}\label{eq:GP_prior_labelsNoise}
    y(\vect{x}) \sim \mathcal{GP}\left(m(\vect{x}),k(\vect{x},\vect{x}^\prime) + \sigma^2\vect{I}\right).
\end{equation}
This GP is used as \textit{prior} for non-parametric Bayesian inference of the latent function. 
Consider the test inputs 
$
    \vect{X}_* = 
\begin{bmatrix}
\vect{x}_{1,*} & \dots & \vect{x}_{n_{test},*}
\end{bmatrix}^\top \! \in \! \mathbb{R}^{n_{test} \times p}.
$
From the definition of a GP, any finite number of samples drawn from the GP are jointly Gaussian. Thus~\cite{williams2006gaussian},
\begin{equation}\label{eq:GP_joint}
    \begin{bmatrix}
    \vect{y} \\ \vect{f}_*
    \end{bmatrix} \!= \!
    \mathcal{N} \left( 
    \begin{bmatrix}
    \bm{\mu} \\
    \bm{\mu}_*
    \end{bmatrix} \! , \!
    \begin{bmatrix}
    \vect{K}(\vect{X},\vect{X})+\sigma^2\vect{I}  
    \! & \! \vect{K}(\vect{X},\vect{X}_*) \\
    \vect{K}(\vect{X}_*,\vect{X}) \! & \!
    \vect{K}(\vect{X}_*,\vect{X}_*)
    \end{bmatrix}
    \right)
\end{equation}
where: $\vect{f}_*=
\begin{bmatrix}
f(\vect{x}_{1,*}) & \dots & f(\vect{x}_{n_{test},*})
\end{bmatrix}^\top \! \in \! \mathbb{R}^{n_{test}}
$;
the \mbox{$i$-th} row of $\bm{\mu}$ and $\bm{\mu}_*$ is $m(\vect{x}_i)$ and $m(\vect{x}_{i,*})$, respectively; 
the \mbox{$(i,j)$-th} entry of $\vect{K}(\vect{X},\vect{X})$ and $\vect{K}(\vect{X},\vect{X}_*)$ is $k(\vect{x}_i,\vect{x}_j)$ and $k(\vect{x}_i,\vect{x}_{j,*})$, respectively.

Making predictions about unobserved values $\vect{x}_{i,*}$ consists in drawing samples from the {\em predictive posterior distribution} of $\vect{f}_*$, given $\mathcal{D}$ and $\vect{X}_*$, that is~\cite{williams2006gaussian}
\begin{equation}\label{eq:GP_posterior}
\begin{split}
    & \vect{f}_*|\mathcal{D},\vect{X}_* \sim \mathcal{N}(\vect{f}_*|\bm{\mu}_{*|\mathcal{D}},\bm{\Sigma}_{*|\mathcal{D}})  \\
    & \bm{\mu}_{*|\mathcal{D}} \! = \! \bm{\mu}_*  \! + \! \vect{K}(\vect{X}_*,\vect{X})\left[ \vect{K}(\vect{X},\vect{X}) \! + \! \sigma^2\vect{I} \right]^{-1}(\vect{y}-\bm{\mu}) \\
    & 
    \begin{split}
        \bm{\Sigma}_{*|\mathcal{D}} 
        & = \vect{K}(\vect{X}_*,\vect{X}_*)\\
        & - \vect{K}(\vect{X}_*,\vect{X})\left[ \vect{K}(\vect{X},\vect{X}) +\sigma^2\vect{I} \right]^{-1} \vect{K}(\vect{X},\vect{X}_*) 
    \end{split}
\end{split}
\end{equation}

\subsubsection{\textbf{GPR-based PAME}}\label{subsubsec:PAME_GPR}

Consider the PAME framework described in Sec. \ref{sec:problem_formulation} and suppose a time-invariant traffic density function, i.e., 
\begin{equation}\label{eq:time_invariant_traffic}
 \mu_{t-T}(x)=\mu_t(x), \; \forall x\in\mathcal{X},\; \forall t. 
\end{equation}
At a fixed time instant $t$, it is possible to employ GPR to learn the PAM function $f(x|\mu_t(x))$. To this aim, we consider $\mathcal{D}_t$ as training dataset and we model the underlying PAM as a GP with zero mean function and Matern covariance function $k(x,x^\prime)$, that is\footnote{The notation $\hat{f}(x)$ is used to distinguish  
the GP model w.r.t. the underlying one, $f(x)$.} 
\begin{equation}\label{eq:PP_GP}
    \hat{f}(x|\mu_t(x)) \sim \mathcal{GP}\left(0,k(x,x^\prime)\right).
\end{equation}

Note that the GPR problem is well-posed~\cite{williams2006gaussian}, according to \eqref{eq:observation_model_gaussian}.

\subsection{Uncertainty-aware active sampling}\label{subsec:selection}

Theoretically, the PAME task can be solved with a standard GPR procedure, as outlined in Sec. \ref{subsec:PAME_GPR}: as a new datapoint is collected at time $t$, it is included in the dataset $\mathcal{D}_{t-T}$, together with the samples coming from the external sources, as in \eqref{eq:dataset_update}; at this point, a new GPR process is performed on $\mathcal{D}_t$. As $|\mathcal{D}_t|$ increases, the GPR converges to an optimal reconstruction of the latent function~\cite{shahriari2015taking,carron2015multi}. Nonetheless, one of the main drawbacks of GPR is the poor scalability properties with respect to the dataset cardinality~\cite{das2018fast,liu2020gaussian,moore2016fast,shahriari2015taking}. In particular, the PAME framework induces a fast intractability in the learning process, since the platform adds to the dataset $N_t+1$ new data at each iteration.

Therefore, we propose an active sensing scheme to select which incoming data (among the $N_t+1$) should be kept in memory (i.e., included in the dataset) and used in the successive GPR iterations.
Motivated by the fast dataset growth and the good estimation properties of GPs with small datasets~\cite{williams2006gaussian}, we suggest to consider only one sample per iteration, 
according to the following selection criterion
$a(x|\mathcal{D}_t)$ (\emph{acquisition function})
\begin{equation}\label{eq:selection}
    x_{t} = \argmax_{x_* \in s_t \cup \{x_i\}_{i=1}^{N_t}} a(x_*|\mathcal{D}_{t-T}),
\end{equation}
where $s_t \cup \{x_i\}_{i=1}^{N_t}$ is the set of query points at time $t$.
The acquisition function $a(\cdot|\mathcal{D}_{t-T})$ is designed over the current GP estimate and quantifies the utility of a query point to produce a more informative posterior distribution; hence, the design of a suitable selection criterion is critical in many applications. In literature a wide variety of acquisition functions are proposed~\cite{shahriari2015taking}: some of them prioritise samples that minimize the uncertainty of the model (or, equivalently, maximize the information gain); some are used to foster a balanced exploration of the latent function domain, while others are designed to focus the estimation quality on specified domain regions; some others take into account the cost associated to the sampling process (useful when query points have non-equal sampling cost). In this work, we use an \textit{uncertainty-aware policy}, that is
\begin{equation}\label{eq:acquisition}
a(x_*|\mathcal{D}_{t-T}) = \sigma_{*,\mathcal{D}_{t-T}}   
\end{equation}
where \eqref{eq:GP_posterior} is applied on the query point $x_*$ and $\sigma_{*,\mathcal{D}_{t-T}}$ is $\bm{\Sigma}_{*,\mathcal{D}_{t-T}}$ with $p=1$.
The predicted variance is a measure of the process entropy~\cite{viseras2016decentralized}; hence, by choosing the position with the highest predicted variance, we sample where the entropy is highest,
that is, at the most informative location. 
In Sec. \ref{sec:numerical} we show that the active sampling module remarkably decreases the computational load of the map learning process; moreover, the acquisition function choice in \eqref{eq:acquisition} is leads to a fast convergence and adaptability behavior.

\subsection{Adaptation to time-varying traffic density}\label{subsec:adaptation}

So far, we have considered a time-invariant traffic density function \eqref{eq:time_invariant_traffic}.
Nonetheless, as stated in Sec. \ref{sec:problem_formulation}, the traffic density is time-varying in real-life scenarios. Consequently, some data in $\mathcal{D}_t$ become obsolete as $\mu(\cdot)$ changes; more formally, 
\begin{equation}\label{eq:obsolete}
    \mathcal{D}_{obsolete} = \{ (x_j,y_j) \in \mathcal{D}_t : \mu_{t-T}(x_j) \neq \mu_t(x_j)  \} \subseteq  \mathcal{D}_t
\end{equation}
is the sub-dataset of obsolete data at time $t$.
Recall that the platform captures information on the current traffic conditions from cloud databases. Thus, it is capable to compute $\mathcal{D}_{obsolete}$ and, therefore, it can remove obsolete data from the current dataset (see Alg. \ref{algo:pipeline}); in this way, the successive GP learning processes are not to contaminated by spurious information. This technique, combined with the multi-source data collection, allows to quickly adapt the parking availability map under dynamic traffic conditions.

\begin{algorithm}[t!]
\small
    \textbf{Initialization:} 
    \begin{itemize}
        \item $s_0 = 0$
        \item $\mathcal{D}_0 = (\emptyset,\emptyset)$
    \end{itemize}
    \While{$s_t \leq L$}{
        - update knowledge on $\mu_t(x)$\\
        - compute $\mathcal{D}_{obsolete}$\\
        \If{$\mathcal{D}_{obsolete} \neq \emptyset$
        }{
            - $\mathcal{D}_{t-T} \leftarrow \mathcal{D}_{t-T} \setminus  \mathcal{D}_{obsolete}$
        }
        - measure $y_t(s_t)$\\
        - receive $\left( \{x_i\}_{i=1}^{N_t} , \{ y_t(x_i) \}_{i=1}^{N_t} \right)$\\
        - solve \eqref{eq:selection} $\rightarrow x_t$\\
        - $\mathcal{D}_t = \mathcal{D}_t \cup x_t$  \\
        - perform GPR on $\mathcal{D}_t \rightarrow \hat{f}(x|\mu_t(x))$ \\
        - $s_{t+T} = s_t + v(\mu_t(x))T$\\
        - $t \leftarrow t+T$
    }
    \textbf{Output:}
    \begin{itemize}
    \item $\hat{f}(x|\mu_t(x)) \sim \mathcal{GP}\left(0,k(x,x^\prime)\right)$
    \end{itemize}
\caption{Online and adaptive PAME.
}
\label{algo:pipeline}
\end{algorithm} 

\section{Numerical results}\label{sec:numerical}



\begin{figure*}[t]
\centering
\includegraphics[width=0.7\textwidth]{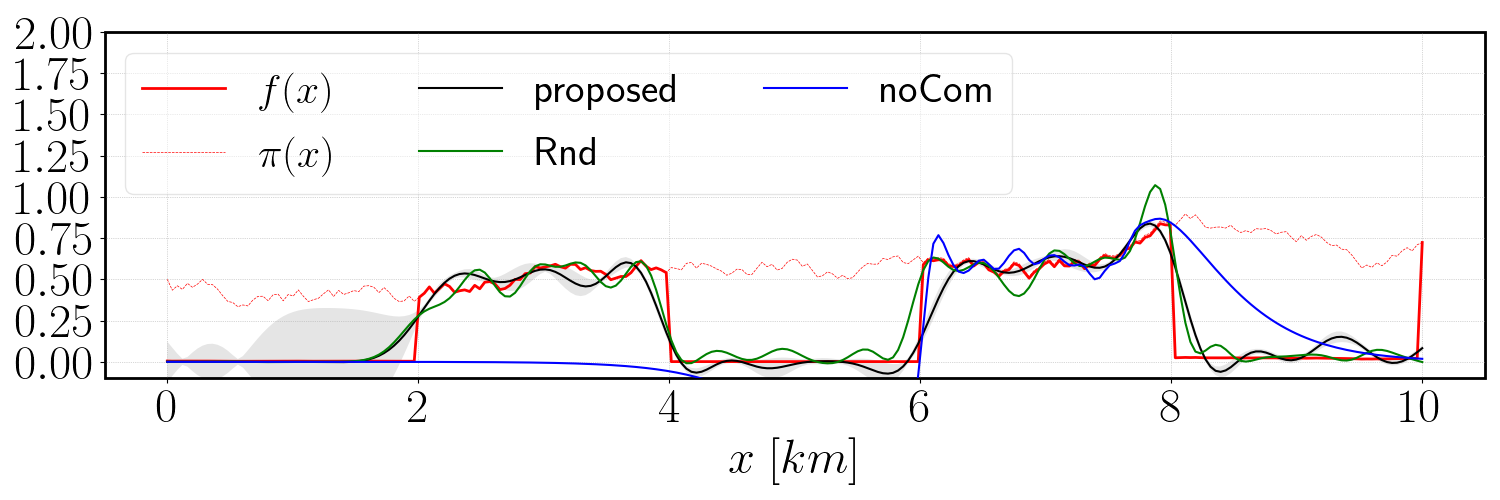}
\caption{Example of synthetic PAM (red) and its a-priori availability function $\pi(x)$ (red dashed). The proposed approach (black) is compared against \textit{Rnd} (green) and \textit{noCom} (blue). The gray shaded area is the uncertainty associated to the GP model obtained with Alg. \ref{algo:pipeline}.}
\label{fig:GP}
\end{figure*}

To evaluate the proposed approach, we consider a Python-based synthetic environment\footnote{https://github.com/luca-varotto/Pparking}.
Sec.~\ref{subsec:params} describes the main setup parameters, as well as the synthetic data generation process. Sec.~\ref{subsec:assessment} defines the metrics used for performance assessment and the baselines considered for comparison. 
The numerical simulation results are discussed in  Sec.~\ref{subsec:discussion_numerical}.


\begin{figure*}[t!]
\centering
\subfigure[]{
\includegraphics[width=0.315\textwidth]{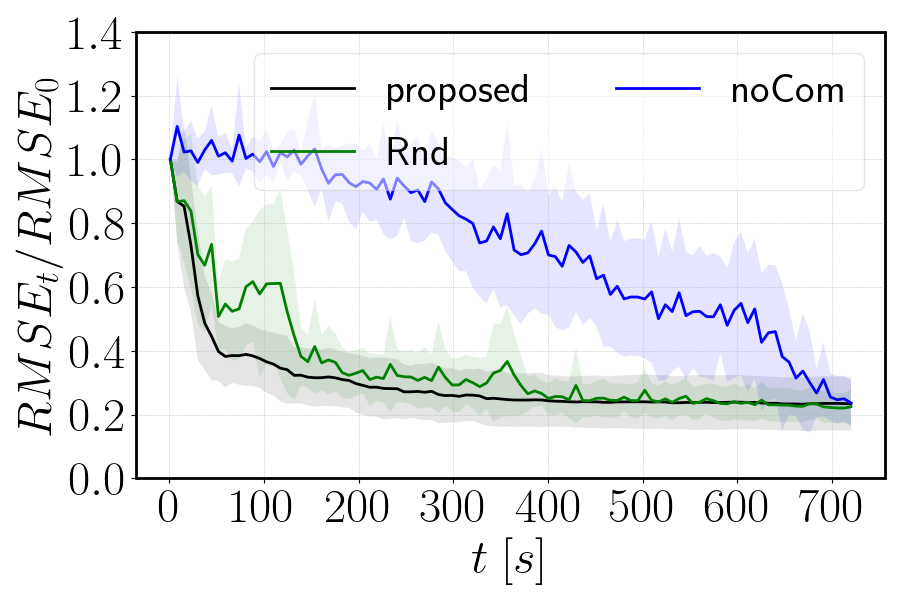}
\label{fig:learning_curve_TimeInvariantTraffic}
}
\subfigure[]{
\includegraphics[width=0.315\textwidth]{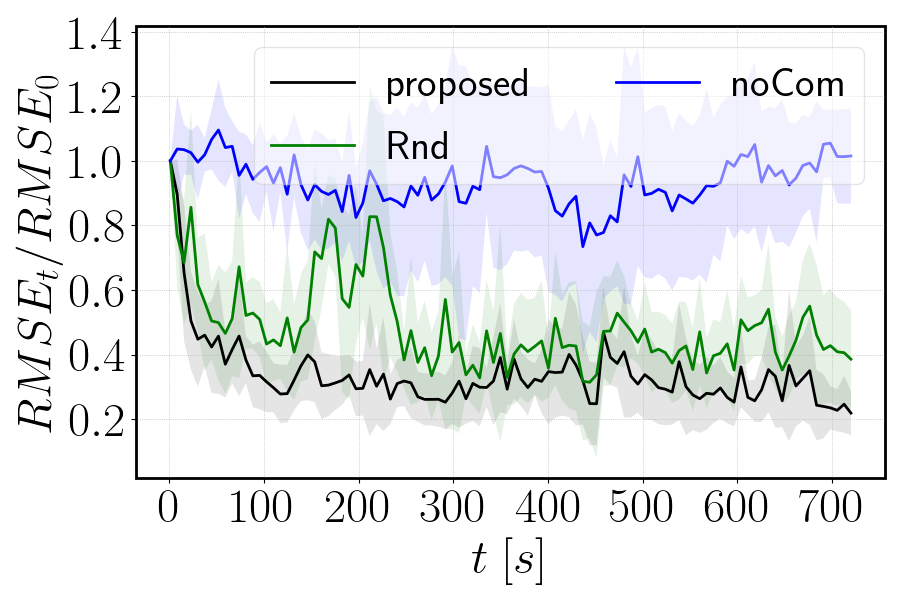}
\label{fig:learning_curve_TimeVariantTraffic}
} 
\subfigure[]{
\includegraphics[width=0.315\textwidth]{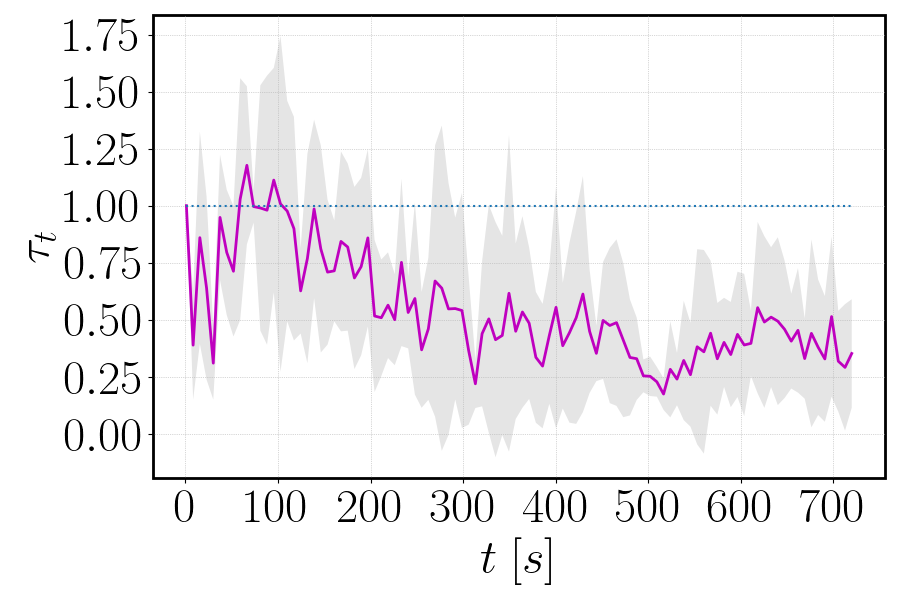}
\label{fig:time_ratio}
} 
\caption{MC numerical results: average (lines) and $68\%$ confidence interval (shaded area). (a) Learning curve in case of time-invariant traffic; (b) learning curve in case of time-varying traffic; (c) processing time ratio in case of time-invariant traffic.}
\label{fig:numerical_results}
\end{figure*}

\subsection{Setup parameters }\label{subsec:params}

In our synthetic environment, the platform is required to travel a length of \mbox{$L=\SI{10}{km}$} with a velocity $v(\mu_t(s_t))$, function of the local traffic density conditions (see Tab. \ref{tab:setup_parameters}). 
The traffic density is a piece-wise constant function where the each constant portion is $\SI{1}{km}$-long and with a value randomly chosen between $0$ and $1$.
The time-variance is simulated through a Bernoulli random variable of parameter $p_{change}=0.2$, which is used to change the value of one of the constant portions (the segment is randomly chosen, as well as its new value).
The a-priori parking availability function is generated according to its definition \eqref{eq:prior}, where $W=\SI{100}{m}$ and the indicator function $\mathds{1}(x)$ is a binary piece-wise constant function where each portion is $\SI{5}{m}$-long (i.e., $D=\SI{5}{m}$) and whose binary value is chosen randomly.   
The number of connected external sources at time $t$ is modeled as $N_t \sim \mathcal{U}(0,10)$. 
Measurements are simulated according to the model \eqref{eq:observation_model_gaussian} with $\sigma=3 \times 10^{-2}$.
For all the remaining relevant parameters, refer to Tab. \ref{tab:setup_parameters}.

\begin{table}[t!]
\vspace{0.3cm}
\centering
\caption{Setup parameters for the MC experiment.}
\label{tab:setup_parameters}
\begin{tabular}{c| c}
\hline
\bf{Parameter} & \bf{Value} \\
\hline
\rowcolor{lightgray}
$\lambda(\mu)$ & $ (1+e^{20(\mu-0.5})^{-1}$\\
$v(\mu)$ & $90e^{-\mu} \SI{}{km/h}$\\
\hline
\rowcolor{lightgray}
$L$ & $\SI{10}{\km}$\\
\hline
$N_{tests}$ & $10$\\
\hline
\rowcolor{lightgray}
$N_t$ & $\sim \mathcal{U}(0,10)$\\
\hline
$p_{change}$ & $0.2$\\
\hline
\rowcolor{lightgray}
$T$ & $\SI{10}{s}$\\
\hline
$D$ & $\SI{5}{m}$\\
\hline
\rowcolor{lightgray}
$W$ & $\SI{100}{m}$\\
\hline
$\sigma$ & $3 \times 10^{-2}$\\
\hline
\end{tabular}
\vspace{-0.3cm}
\end{table}

\subsection{Performance assessment}\label{subsec:assessment}

To capture the simulation and performance variability, numerical evaluation is performed through a Monte Carlo (MC) experiment, composed by $N_{tests} = 10$ tests.

\subsubsection{\textbf{Baselines}}\label{subsubsec:baselines}
The following baselines are considered for comparison.
\begin{itemize}
    \item \textit{NoCom}: the platform is a non-connected vehicle; hence, it relies on its measurements only (i.e., $s_t$).
    \item \textit{NoSel}: the platform is a connected vehicle, but the active sampling module is not applied (i.e., $\mathcal{D}_t$ is updated with all $N_t$ available data).
    \item \textit{Rnd}: the platform is a connected vehicle and the active sampling module uses a random acquisition function, namely \eqref{eq:selection} is substituted by a random choice in $s_t \cup \{x_i\}_{i=1}^{N_t}$.
\end{itemize}

Fig. \ref{fig:GP} shows a realization of $\pi(x)$, with the corresponding PAM function to be estimated, according to the synthetic generation procedure detailed in \mbox{Sec. \ref{subsec:params}}. As can be seen, \textit{noCom} estimates correctly only on a small neighbourhood of the current platform position (\mbox{$s_t= \SI{7.5}{km}$}). In principle, its GP model should be accurate for any $x \leq s_t$, because its training dataset is composed by $\{s_k\}_{k=0}^t$, where \mbox{$s_k < s_t,\; k<t$}. Nevertheless, many datapoints become obsolete over time, due to the dynamic traffic conditions; thus, the accuracy remains good only locally (i.e., where samples are recent). To sum up, \textit{noCom} has poor adaptivity capabilities, since obsolete data can not be replaced by more up to date samples; for this reason, it can not accurately estimate the latent function over the entire domain $\mathcal{X}$.

\subsubsection{\textbf{Performance metrics}}\label{subsubsec:metrics}
The following performance indexes are computed from the MC simulation. 
\begin{itemize}
    \item \textit{Learning curve}: 
    we evaluate the estimation quality at time $t$ through the Root Mean Squared Error between the underlying function $f(\cdot|\mu_t(\cdot))$ and its estimated version $\hat{f}(\cdot|\mu_t(\cdot))$, i.e.
    \begin{equation}\label{eq:rmse}
        RMSE_t = \sqrt{ \frac{1}{L} \int_{0}^{L} \left[ \hat{f}(x|\mu_t(x)) - f(x|\mu_t(x)) \right]^2 dx}.
    \end{equation}
The learning curve is defined as $RMSE_t/RMSE_0$, where $RMSE_0$ is computed before any measurement is collected (i.e., during model initialization) and it is equal for all baselines, since we fixed the initial model parameters.
    \item \textit{Processing time ratio}: it is computed as 
    \begin{equation}\label{eq:proc_ratio}
        \tau_t = \frac{T_{C,t}^{(proposed)}}{T_{C,t}^{(\textit{noSel})}}
    \end{equation}
    where $T_{C,t}^{(proposed)}$ and $T_{C,t}^{(\textit{noSel})}$ are the computation time for the proposed algorithm and \textit{noSel}, respectively; both are referred to the  an entire GPR iteration: model training and prediction. 
\end{itemize}

\subsection{Discussion}\label{subsec:discussion_numerical}

\subsubsection{\textbf{Time-invariant traffic}}\label{subsubsec:time_invariant} 

We first consider the simplified scenario in which the traffic density is time-invariant; this allows to analyze the asymptotic estimation properties of the methods under comparison, since the function to be learnt does not change in time. As depicted in Fig. \ref{fig:learning_curve_TimeInvariantTraffic}, the proposed approach has the fastest and smoothest convergence; more specifically, during the entire MC experiment, the RMSE ratio between the proposed and \textit{Rnd} is smaller than $1$ for the $66\%$ of the times, and this rate become $96\%$ if we compare with \textit{noCom} (the $4\%$ is due to small oscillations at steady-state). From these results, the benefit of an uncertainty-aware acquisition function becomes clear: it drives the data collection process towards the highest information gain, ensuring fast and high quality mapping performance. Interestingly, the absence of any communication with the external sources (i.e., \textit{noCom}) induces the slowest convergence; in particular, the platform does not have spatial predictive capabilities and needs to reach the end of the path to have enough information to accurately reconstruct the latent function.


\subsubsection{\textbf{Impact of uncertainty-aware sampling}}\label{subsubsec:time_variant}

The main benefits of the proposed strategy arise when the traffic conditions are time-varying. The local information exploited by \textit{NoCom} and its poor adaptivity capabilities (see Sec. \ref{subsubsec:baselines}), make this algorithm not suitable for dynamic traffic conditions, as shown in Fig. \ref{fig:learning_curve_TimeVariantTraffic}. Therefore, in this scenario it is fundamental to exploit information coming from external sources. In particular, the fast adaptation capabilities of multi-source data collection, coupled with an active sampling strategy, guarantee satisfactory estimation performance. As described in Sec. \ref{subsec:adaptation} and in Alg. \ref{algo:pipeline}, when traffic density variations are detected, the corresponding obsolete datapoints are removed from the dataset. Hence, the GP model uncertainty associated to the removed samples increases; consequently, the policy in \eqref{eq:acquisition} will prioritise new query points coming from those regions where any traffic change has been detected; this leads to a faster recovery of the estimation process w.r.t. to a random sampling process. Indeed, Fig. \ref{fig:learning_curve_TimeVariantTraffic} shows a remarkable performance difference between the proposed algorithm and the baseline \textit{Rnd}; in fact, the RMSE ratio of \textit{Rnd} is $80\%$ of the times larger than ours ($14\%$ more than the time-invariant case).


\subsubsection{\textbf{Impact of data selection}}\label{subsubsec:data_selection}
Fig.  \ref{fig:time_ratio} shows the processing time ratio and highlights the benefit of performing data selection when dealing with GPR and large scale data acquisition processes. The time required to fit the GP model and to estimate the underlying function with the proposed approach is $83\%$ of the times smaller than a non selective strategy. Notably, the computational gap between the two methods increases (on average) as the dataset increases, as expected from \eqref{eq:dataset_cardinality}.


\section{Conclusion }\label{sec:conclusion}

This work proposes an online and adaptive learning technique for estimating the on-street parking probability in the framework of connected vehicles. 
The parking availability map is incrementally learnt via GPR in a multi-source data collection scenario. To prevent computational and storage intractability issues, an information-driven active sensing module is applied to select incoming data. Numerical results validate the proposed approach in terms of adaptation capabilities, learning speed, and  execution time.

Future work will be devoted to the extension on multi-platform cooperative scenarios.
{\renewcommand{\baselinestretch}{0.988}
\bibliographystyle{IEEEtran}
\bibliography{IEEEfull,References}

\begin{thebibliography}{10}
\providecommand{\url}[1]{#1}
\csname url@samestyle\endcsname
\providecommand{\newblock}{\relax}
\providecommand{\bibinfo}[2]{#2}
\providecommand{\BIBentrySTDinterwordspacing}{\spaceskip=0pt\relax}
\providecommand{\BIBentryALTinterwordstretchfactor}{4}
\providecommand{\BIBentryALTinterwordspacing}{\spaceskip=\fontdimen2\font plus
\BIBentryALTinterwordstretchfactor\fontdimen3\font minus
  \fontdimen4\font\relax}
\providecommand{\BIBforeignlanguage}[2]{{%
\expandafter\ifx\csname l@#1\endcsname\relax
\typeout{** WARNING: IEEEtran.bst: No hyphenation pattern has been}%
\typeout{** loaded for the language `#1'. Using the pattern for}%
\typeout{** the default language instead.}%
\else
\language=\csname l@#1\endcsname
\fi
#2}}
\providecommand{\BIBdecl}{\relax}
\BIBdecl

\bibitem{sae2014taxonomy}
{SAE On-Road Automated Vehicle Standards Committee and others}, ``Taxonomy and
  definitions for terms related to on-road motor vehicle automated driving
  systems,'' \emph{SAE Standard J}, vol. 3016, pp. 1--16, 2014.

\bibitem{joy2017internet}
J.~Joy and M.~Gerla, ``Internet of vehicles and autonomous connected
  car-privacy and security issues,'' in \emph{2017 26th International
  Conference on Computer Communication and Networks (ICCCN)}.\hskip 1em plus
  0.5em minus 0.4em\relax IEEE, 2017, pp. 1--9.

\bibitem{gopalswamy2018infrastructure}
S.~Gopalswamy and S.~Rathinam, ``Infrastructure enabled autonomy: A distributed
  intelligence architecture for autonomous vehicles,'' in \emph{2018 IEEE
  Intelligent Vehicles Symposium (IV)}.\hskip 1em plus 0.5em minus 0.4em\relax
  IEEE, 2018, pp. 986--992.

\bibitem{uhlemann2015introducing}
E.~Uhlemann, ``Introducing connected vehicles [connected vehicles],''
  \emph{IEEE Vehicular Technology Magazine}, vol.~10, no.~1, pp. 23--31, 2015.

\bibitem{uhlemann2018time}
------, ``Time for autonomous vehicles to connect [connected vehicles],''
  \emph{IEEE vehicular technology magazine}, vol.~13, no.~3, pp. 10--13, 2018.

\bibitem{wong2021mapping}
K.~Wong, Y.~Gu, and S.~Kamijo, ``Mapping for autonomous driving: Opportunities
  and challenges,'' \emph{IEEE Intelligent Transportation Systems Magazine},
  vol.~13, no.~1, 2021.

\bibitem{siegel2017survey}
J.~E. Siegel, D.~C. Erb, and S.~E. Sarma, ``A survey of the connected vehicle
  landscape—architectures, enabling technologies, applications, and
  development areas,'' \emph{IEEE Transactions on Intelligent Transportation
  Systems}, vol.~19, no.~8, pp. 2391--2406, 2017.

\bibitem{varotto2021probabilistic}
L.~Varotto, A.~Cenedese, and A.~Cavallaro, ``Probabilistic radio-visual active
  sensing for search and tracking,'' 2021.

\bibitem{varotto2021transmitter}
L.~Varotto and A.~Cenedese, ``Transmitter discovery through radio-visual
  probabilistic active sensing,'' 2021.

\bibitem{talebpour2016influence}
A.~Talebpour and H.~S. Mahmassani, ``Influence of connected and autonomous
  vehicles on traffic flow stability and throughput,'' \emph{Transportation
  Research Part C: Emerging Technologies}, vol.~71, pp. 143--163, 2016.

\bibitem{lin2017survey}
T.~Lin, H.~Rivano, and F.~Le~Mou{\"e}l, ``A survey of smart parking
  solutions,'' \emph{IEEE Transactions on Intelligent Transportation Systems},
  vol.~18, no.~12, pp. 3229--3253, 2017.

\bibitem{bischoff2018autonomous}
J.~Bischoff, M.~Maciejewski, T.~Schlenther, and K.~Nagel, ``Autonomous vehicles
  and their impact on parking search,'' \emph{IEEE Intelligent Transportation
  Systems Magazine}, vol.~11, no.~4, pp. 19--27, 2018.

\bibitem{westfechtel2018parking}
T.~Westfechtel, K.~Ohno, N.~Mizuno, R.~Hamada, S.~Kojima, and S.~Tadokoro,
  ``Parking spot estimation and mapping method for mobile robots,'' \emph{IEEE
  Robotics and Automation Letters}, vol.~3, no.~4, pp. 3371--3378, 2018.

\bibitem{li2018collaborative}
B.~Li, L.~Yang, J.~Xiao, R.~Valde, M.~Wrenn, and J.~Leflar, ``Collaborative
  mapping and autonomous parking for multi-story parking garage,'' \emph{IEEE
  Transactions on Intelligent Transportation Systems}, vol.~19, no.~5, pp.
  1629--1639, 2018.

\bibitem{mathur2010parknet}
S.~Mathur, T.~Jin, N.~Kasturirangan, J.~Chandrasekaran, W.~Xue, M.~Gruteser,
  and W.~Trappe, ``Parknet: drive-by sensing of road-side parking statistics,''
  in \emph{Proceedings of the 8th international conference on Mobile systems,
  applications, and services}, 2010, pp. 123--136.

\bibitem{o2009contextual}
S.~O'Callaghan, F.~T. Ramos, and H.~Durrant-Whyte, ``Contextual occupancy maps
  using gaussian processes,'' in \emph{2009 IEEE International Conference on
  Robotics and Automation}.\hskip 1em plus 0.5em minus 0.4em\relax IEEE, 2009,
  pp. 1054--1060.

\bibitem{senanayake2017learning}
R.~Senanayake, S.~O'Callaghan, and F.~Ramos, ``Learning highly dynamic
  environments with stochastic variational inference,'' in \emph{2017 IEEE
  International Conference on Robotics and Automation (ICRA)}.\hskip 1em plus
  0.5em minus 0.4em\relax IEEE, 2017, pp. 2532--2539.

\bibitem{rajabioun2015street}
T.~Rajabioun and P.~A. Ioannou, ``On-street and off-street parking availability
  prediction using multivariate spatiotemporal models,'' \emph{IEEE
  Transactions on Intelligent Transportation Systems}, vol.~16, no.~5, pp.
  2913--2924, 2015.

\bibitem{tavafoghi2019queuing}
H.~Tavafoghi, K.~Poolla, and P.~Varaiya, ``A queuing approach to parking:
  Modeling, verification, and prediction,'' \emph{arXiv preprint
  arXiv:1908.11479}, 2019.

\bibitem{williams2006gaussian}
C.~K. Williams and C.~E. Rasmussen, \emph{Gaussian processes for machine
  learning}.\hskip 1em plus 0.5em minus 0.4em\relax MIT press Cambridge, MA,
  2006.

\bibitem{krasniqi2016use}
X.~Krasniqi and E.~Hajrizi, ``Use of iot technology to drive the automotive
  industry from connected to full autonomous vehicles,''
  \emph{IFAC-PapersOnLine}, vol.~49, no.~29, pp. 269--274, 2016.

\bibitem{aryandoust2019city}
A.~Aryandoust, O.~van Vliet, and A.~Patt, ``City-scale car traffic and parking
  density maps from uber movement travel time data,'' \emph{Scientific data},
  vol.~6, no.~1, pp. 1--18, 2019.

\bibitem{cao2015system}
J.~Cao and M.~Menendez, ``System dynamics of urban traffic based on its
  parking-related-states,'' \emph{Transportation Research Part B:
  Methodological}, vol.~81, pp. 718--736, 2015.

\bibitem{varotto2021wakeUp}
L.~Varotto and A.~Cenedese, ``Probabilistic rf-assisted camera wake-up through
  self-supervised gaussian process regression,'' 2021.

\bibitem{das2018fast}
S.~Das, S.~Roy, and R.~Sambasivan, ``Fast gaussian process regression for big
  data,'' \emph{Big data research}, vol.~14, pp. 12--26, 2018.

\bibitem{liu2020gaussian}
H.~Liu, Y.-S. Ong, X.~Shen, and J.~Cai, ``When gaussian process meets big data:
  A review of scalable gps,'' \emph{IEEE transactions on neural networks and
  learning systems}, vol.~31, no.~11, pp. 4405--4423, 2020.

\bibitem{moore2016fast}
C.~J. Moore, A.~J. Chua, C.~P. Berry, and J.~R. Gair, ``Fast methods for
  training gaussian processes on large datasets,'' \emph{Royal Society open
  science}, vol.~3, no.~5, p. 160125, 2016.

\bibitem{shahriari2015taking}
B.~Shahriari, K.~Swersky, Z.~Wang, R.~P. Adams, and N.~De~Freitas, ``Taking the
  human out of the loop: A review of bayesian optimization,'' \emph{Proceedings
  of the IEEE}, vol. 104, no.~1, pp. 148--175, 2015.

\bibitem{carron2015multi}
A.~Carron, M.~Todescato, R.~Carli, L.~Schenato, and G.~Pillonetto,
  ``Multi-agents adaptive estimation and coverage control using gaussian
  regression,'' in \emph{2015 European Control Conference (ECC)}.\hskip 1em
  plus 0.5em minus 0.4em\relax IEEE, 2015, pp. 2490--2495.

\bibitem{viseras2016decentralized}
A.~Viseras, T.~Wiedemann, C.~Manss, L.~Magel, J.~Mueller, D.~Shutin, and
  L.~Merino, ``Decentralized multi-agent exploration with online-learning of
  gaussian processes,'' in \emph{2016 IEEE International Conference on Robotics
  and Automation (ICRA)}.\hskip 1em plus 0.5em minus 0.4em\relax IEEE, 2016,
  pp. 4222--4229.

\end{thebibliography}
}
\vfill
\end{document}